\documentclass[aps,showpacs,preprintnumbers,amsmath,amssymb,nofootinbib]{revtex4}
\begin{document}
\title{Stochastic gravitoelectromagnetic inflation}
\author{$^{1}$Jos\'e Edgar Madriz Aguilar\footnote{
E-mail address: jemadriz@fisica.ufpb.br} and
$^2$Mauricio Bellini\footnote{E-mail address: mbellini@mdp.edu.ar}}

\address{$^1$ Departamento de F\'{\i}sica, Universidade Federal da Para\'{\i}ba. C. Postal 5008, Jo\~{a}o Pessoa, PB 
58059-970 Brazil.\\
$^2$ Consejo Nacional de Investigaciones
Cient\'{\i}ficas y T\'ecnicas (CONICET) and Departamento de
F\'{\i}sica, Facultad de Ciencias Exactas y Naturales, Universidad
Nacional de Mar del Plata, Funes 3350, (7600) Mar del Plata,
Argentina.}

\vskip .5cm

\begin{abstract}
Gravitoelectromagnetic inflation was recently introduced
to describe, in an unified manner,
electromagnetic, gravitatory and
inflaton fields in the early (accelerated) inflationary
universe from a 5D vacuum state.
In this paper, we study a stochastic treatment for
the gravitoelectromagnetic components $A_B=(A_{\mu},\varphi)$, on
cosmological scales. We focus our study
on the seed magnetic fields on super Hubble scales, which could
play an important role in large scale structure
formation of the universe.
\end{abstract}
\maketitle \vskip .2cm \noindent
\vskip 1cm
\section{Introduction}

Most of the cosmologists believe that our universe has
experienced an early period of accelerated expansion, called
inflation\cite{star,guth,linde}.
Inflation provides a mechanism that explains the origin of the large scale structure formation process.
In this mechanism quantum fluctuations  of the inflaton field that were stretched beyond the horizon become classical 
perturbations.
In the stochastic approach of inflation the dynamics
of the quantum to classical transition is effectively described
by a classical noise\cite{habib,prd96}.
In this approach the scalar field is coarse grained. Thus, the
field modes are splited into a subhorizon and a superhorizon parts.
The superhorizon part, which describes the fluctuations on cosmological
scales, constitutes the classical and homogeneous coarse-grained field driven
by the stochastic noise due to the subhorizon modes that are
leaving the horizon. As the inflationary universe expands rapidly, more
and more short wavelength modes are stretched beyond the horizon
and thus the number of degrees of freedom of the coarse-grained field is being increased.
This phenomena is viewed by this field as a noise.
The dynamics of this field
can be described by a second order stochastic equation, which can be
rewritten as two first order (Langevin) equations. The stochastic
properties of the relevant noise depend of the window function that
separates the sub and superhorizon modes\cite{npb}.

On the other hand, the possible existence, strength and structure
of magnetic fields in the intergalactic plane, within the Local
Superclusted, has been scrutinized recently\cite{ocho}. If the
seed of these magnetic fields was originated during inflation, the
study of its origin and evolution in this epoch should be very
important in cosmology\cite{giovannini}. The origin of the
primordial magnetic fields has been subject of a great amount of
research\cite{mag}. These concepts can be extended and worked in
the gravitoelectromagnetic inflation formalism recently introduced
in \cite{grav1}. Gravitoelectromagnetic inflation was developed
with the aim to describe, in an unified manner, the inflaton,
gravitatory and electromagnetic fields during inflation. The
formalism has the adventage that all the 4D sources has a
geometrical origin and can explain the origin of seed of magnetic
fields on cosmological scales observed today .
Gravitoelectromagnetic inflation was constructed from a 5D vacuum
state on a $R^{A}\,_{BCD}=0$ globally flat metric. As in all Space
Time Matter (STM) models\cite{.}, the 4D sources are geometrically
induced when we take a foliation on the fifth coordinate which is
spacelike and noncompact. There is a main difference between STM
and Brane-World (BW)\cite{...} formalisms. In the STM theory of
gravity we do not need to insert any matter into the 5D manifold
by hand, as is commonly
done in the BW formalism. The 5D metrics used in the STM theory are exact solutions of the
5D field equations in apparent vacuum. The interesting thing here, is that
matter appears in four dimensions without any dimensional compactification,
but induced by the 5D vacuum conditions.

The aim of this paper is to develop a stochastic treatment
for the components $A_{B}=(A^{\mu},\varphi)$ on cosmological scales,
to be able to make a comparison with the results obtained
in a previous paper\cite{grav1}.

\section{ 5D Formalism}

In order to describe a 5D vacuum, we consider
the 5D canonical metric\cite{lb}
\begin{equation}\label{eq1}
dS^{2}=\psi^{2}dN^{2}-\psi^{2}e^{2N}dr^{2}-d\psi^{2},
\end{equation}
where $dr^{2}=dx^{2}+dy^{2}+dz^{2}$. In this line
element the coordinates $(N,r)$ are dimensionless and the
fifth one $\psi$ has spatial units. This metric
describes a
5D flat manifold in apparent vacuum $G_{AB}=0$\footnote{In our conventions,
capital Latin indices run from 0 to 4 and greek indices from 0 to 3.} and
satisfies $R^{A}\,_{BCD}=0$.
To describe an electromagnetic field
and  neutral matter on this background, we consider the action
\begin{equation}\label{eq2}
I=\int d^{4}x \  d\psi\sqrt{\left|\frac{^{(5)}g}{^{(5)}g_0}\right|}
\left[\frac{^{(5)}R}{16\pi {\rm G}} + ^{(5)}{\cal L}(A_{B},A_{C;B})\right],
\end{equation}
for a vector potencial with components
$A_{B}=(A_{\mu},\varphi)$, which
are minimally coupled to gravity.
Here $^{(5)}R$ is the 5D Ricci scalar, which is zero for the metric
(\ref{eq1}) and $\left| ^{(5)} g_0\right|=\psi^8_0$ is a
constant of dimensionalization determined by
$\left| ^{(5)} g\right|=\psi^8 e^{6 N}$ evaluated at $\psi=\psi_0$ and
$N=0$. We propose a 5D lagrangian density in (\ref{eq2})
\begin{equation}\label{eq3}
^{(5)}{\cal L}(A_{B},A_{B;C})=-\frac{1}{4}Q_{BC}Q^{BC}
\end{equation}
where we define the tensor field $Q_{BC}=F_{BC}
+\gamma g_{BC}\left(A^{D}\,_{;D}\right)$,
with $\gamma =\sqrt{(2\lambda / 5)}$ and $F_{BC}=A_{C;B}-A_{B;C}
=-F_{CB}$, being $(;)$ the covariant derivative. The lagrangian
density (\ref{eq3}) can also be expressed as
\begin{equation}\label{ex2}
 ^{(5)}{\cal L}(A_{B},A_{B;C})=-\frac{1}{4}F_{BC}F^{BC}
 -\frac{\lambda}{2}\left(A^{D}\,_{;D}\right)^{2},
 \end{equation}
the last term being a ``gauge-fixing'' term. The
5D-dynamics field equations in a Lagrange formalism
leads to
\begin{equation}\label{eq4}
A^{B}\,_{;D}\,^{;D}-(1-\lambda)A^{C}\,_{;C}\,^{;B}=0.
\end{equation}
Working in the Feynman gauge $(\lambda =1)$, equation (\ref{eq4}) yields
\begin{equation}\label{eq6}
\frac{1}{\sqrt{\left|^{(5)}g\right|}}\frac{\partial}{\partial 
x^{C}}\left[\sqrt{\left|^{(5)}g\right|}\,g^{DC}A^{B}\,_{,D}\right]=0,
\end{equation}
considering all the time that  $A^{B}=(A^{\mu},-\varphi)$. Equation
(\ref{eq6}) is a massless Klein-Gordon-like equation for $A^{B}$
and represents the analogous of
Maxwell's equations in a 5D manifold in an apparent vacuum.
The commutators for $A^C$ and
$\bar{\Pi}^{B}=\frac{\partial {\cal L}}{\partial (A_{B ,N})}=
F^{BN} - g^{B N}A^{C}\,_{;C}$ are given by
\begin{eqnarray}\label{dif1}
\left[A^{C}(N,\vec{r},\psi),\bar{\Pi}^{B}(N,\vec{r'},\psi ')\right]
&=& ig^{CB}
g^{NN} \left|\frac{^{(5)}g_{0}}{^{(5)}g}
\right|\delta^{(3)}\left(\vec{r}-\vec{r'}\right)\delta\left(
\psi - \psi '\right),
\\ \label{dif2}
\left[A_{C}(N,\vec{r},\psi),A_{B}(N,\vec{r'},\psi ')\right]
&=&\left[\bar{\Pi}_{C}(N,\vec{r},\psi),\bar{\Pi}_{B}
(N,\vec{r'},\psi ')\right]=0.
\end{eqnarray}
Here $\bar{\Pi}^N = - g^{NN} \left(A^{C}\,_{;C}\right)$
and $\left|\frac{^{(5)}g_{0}}{^{(5)}g}\right|$ is the inverse of the
normalized volume of the manifold (\ref{eq1}).
From equations (\ref{dif1}) and (\ref{dif2}), we obtain
\begin{equation}
\left[ A_C(N,\vec r,\psi), A_{B;N}(N, \vec{r'},\psi')\right] =
- i \  g_{BC} \left|\frac{^{(5)} g_0}{^{(5)} g}\right|
\delta^{(3)}\left(\vec r - \vec{r'}\right) \  \delta\left(
\psi - \psi'\right).
\end{equation}
Using equations (\ref{eq1}) and (\ref{eq6}), the equation of motion for the
electromagnetic 4-vector potential $A^{\mu}$, is given by
(the overstar denotes
derivative with respect to $N$)
\begin{equation}\label{eq9}
\stackrel{\star \star}{A^{\mu}}+3\stackrel{\star}{A^{\mu}}
-e^{-2N}\nabla _{r}^{2}A^{\mu}-\left[4\psi \frac{\partial A^{\mu}}{\partial
\psi}+\psi^{2}\frac{\partial^2
A^{\mu}}{\partial \psi^{2}}\right]=0,
\end{equation}
where
\begin{eqnarray}
A^{\mu}(N,\vec r,\psi) & = & e^{-3N/2} \left(\frac{\psi_0}{\psi}\right)^2
{\cal A}^{\mu}(N,\vec r,\psi), \qquad {\rm being} \label{chu1} \\
&& {\cal A}^{\mu}(N, \vec r,\psi) = \frac{1}{(2\pi)^{3/2}}
{\large\int} d^3 k_r {\large\int} dk_{\psi} \sum _{\lambda=0}^{3}
\epsilon^{\beta}_{(\lambda)} \left[
a^{(\beta)}_{k_r k_{\psi}}
e^{i [\vec{k_r}.\vec r + k_{\psi} \psi]}
\tilde{Q}_{k_r k_{\psi}}(N,\psi)\right.
\nonumber \\
& + & \left. {a^{(\beta)}}^{\dagger}_{k_r k_{\psi}}
e^{-i [\vec{k_r}.\vec r + k_{\psi} \psi]} \tilde{Q}^*_{k_r k_{\psi}}(N,\psi)
\right], \label{chu2}
\end{eqnarray}
such that $\tilde{Q}_{k_r k_{\psi}}(N,\psi)=e^{-i k_{\psi}\psi}
Q_{k_r k_{\psi}}(N)$ and thus $A^{\mu}(N,\vec r,\psi)= e^{-3N/2}
\left(\psi_0/\psi\right)^2 {\cal A}^{\mu}(N,\vec r)$. Similarly,
for $\varphi$, we have
\begin{equation}\label{eq10}
\stackrel{\star\star}{\varphi} +3\stackrel{\star}{\varphi}-e^{-2N}
\nabla _{r}^{2}\varphi - \left[4\psi\frac{\partial\varphi}{\partial \psi}
+\psi^{2}\frac{\partial ^{2}\varphi}{\partial\psi^{2}}\right]=0.
\end{equation}
According to the 5D flat geometry ($R^A_{BCD}=0$) here used in the
action (\ref{eq2}), the 5D vacuum is described by the Einstein
equations $G_{AB} = 8\pi G T_{AB}=0$, being $T^A_B$ the
energy-momentum tensor. Hence, the only valid solutions for
$A^{\mu}$ and $\varphi$ on the metric (\ref{eq1}) should describe
absence of matter: $A^C_{;C}=0$ (with $A^0=0$ and $A^4=0$).
Furthermore, using (\ref{dif1})
the commutator between $\varphi$ and $\stackrel{\star}{\varphi}$
becomes
\begin{equation}\label{eq12}
\left[\varphi (N,\vec{r},\psi),\stackrel{\star}{\varphi}(N,\vec{r'},
\psi ')\right]
=i\left|\frac{^{(5)}g_{0}}{^{(5)}g}\right|
\delta^{(3)}\left(\vec{r}-\vec{r'}\right)\delta\left(\psi-\psi '\right).
\end{equation}
which is the same expression obtained in \cite{MB1}.

\section{Effective 4D dynamics.}

Considering the coordinate transformations
\begin{equation}\label{eq23}
t=\psi _{0}N,\quad R=r\psi _{0},\quad \psi=\psi,
\end{equation}
equation (\ref{eq1}) takes the form
\begin{equation}\label{eq24}
dS^{2}=\left(\frac{\psi}{\psi _0}\right)^{2}
\left[dt^{2}-e^{2t/\psi _0}dR^{2}\right]-d\psi ^{2} ,
\end{equation}
which is the Ponce Leon metric that describes a 3D spatially
flat, isotropic and homogeneous
extension to 5D of a Friedmann Robertson Walker (FRW)
line element in a de Sitter expansion.
Here $t$ is the cosmic time and $dR^{2}=dX^{2}+dY^{2}+dZ^{2}$.\\

\subsection{The effective 4D electromagnetic field $A^{\mu}$.}

Now, we can take the foliation $\psi=\psi _0$ in
(\ref{eq24}), such that we obtain the effective 4D metric for $R=\psi_0 r$
\begin{equation}\label{eq25}
dS^{2}\rightarrow ds^{2}=dt^{2}-e^{2H_{0}t}dR^{2},
\end{equation}
which describes a 3D spatially flat, isotropic and homogeneous
de Sitter expanding Universe
with constant Hubble parameter
$H_0=1/\psi _0$ and a 4D scalar curvature $^{(4)}{\cal R}=12H^{2}_0$.
The effective 4D action on the effective 4D metric (\ref{eq25}) is
($\mu,\nu =0,1,2,3$)
\begin{equation}\label{act1}
^{(4)} I = \left.{\Large\int} d^4 x \sqrt{\left|\frac{^{(4)} g}{^{(4)} g_0}
\right|} \left[ \frac{^{(4)} {\cal R}}{16\pi G} - \frac{1}{4}
\left[ Q_{\mu\nu} Q^{\mu\nu} + Q_{\psi\psi} Q^{\psi\psi} \right]\right]
\right|_{\psi=\psi_0=H^{-1}_0},
\end{equation}
where the additional term $\left. (1/ 4) Q_{\psi\psi}
Q^{\psi\psi}\right|_{\psi=\psi_0=H^{-1}_0}$ can be identified as
an effective 4D potential. This potential has a geometrical origin
and can take different representations in different frames. In our
case, the observer is in the frame described by the velocities
$U^{\psi}=U^r=0$ (and hence $U^R=0$) and $U^t=1$. The effective
equation of motion for $A^{\mu}(t, \vec R, \psi=\psi_0) \equiv
A^{\mu}(t, \vec R)$\cite{grav1}
\begin{equation}\label{eq26}
\ddot{A}^{\mu}+3H_{0}\dot{A}^{\mu}-
e^{-2H_{0}t}\nabla _{R}^{2}A^{\mu}-H_{0}^{2}\left.
\left[4\psi\frac{\partial A^{\mu}}{\partial \psi}+\psi^{2}\frac{
\partial^{2}A^{\mu}}{\partial \psi ^{2}}
\right]\right|_{\psi=H_{0}^{-1}}=0,
\end{equation}
where, using equations (\ref{chu1}) and (\ref{chu2}), we obtain
$\left. 4\psi (\partial A^{\mu} / \partial\psi) + \psi^2
(\partial^2 A^{\mu}/ \partial\psi^2)\right|_{\psi=H^{-1}_0} =
\left.-2 A^{\mu}\right|_{\psi=H^{-1}_0}$. Here,
$A^{\mu}=A^{\mu}(t,\vec{R})$ is the effective 4D electromagnetic
field induced onto the hypersurface $\psi=H_{0}^{-1}$ and the
overdot denotes the derivative with respect to time. Note that the
last term between brackets in eq. (\ref{eq26}) acts as an induced
electromagnetic potential derived with respect $A^{\mu}$. This
term is the analogous to $V'(\varphi)$ in the case of an
inflationary scalar field as used in \cite{madbe}, and in our case
the dynamics of the component $A^{4} \equiv -\varphi$ is described
by
\begin{equation}\label{eq26'}
\ddot{\varphi}+3H_{0}\dot{\varphi}-
e^{-2H_{0}t}\nabla _{R}^{2}\varphi-H_{0}^{2}\left.
\left[4\psi\frac{\partial
\varphi}{\partial \psi}+\psi^{2}\frac{\partial ^{2}\varphi}{\partial
\psi ^{2}}
\right]\right|_{\psi=H_{0}^{-1}}=0.
\end{equation}
On the other hand, transforming the field $A^\mu$ through the
expression $A^{\mu}(t,\vec{R})$ $=\left. e^{-3N/2}
\left(\psi_0/\psi\right)^2 {\cal A}^{\mu}(N,\vec r,\psi)
\right|_{N=H_0 t, r=R/\psi_0,\psi_0=H^{-1}_0} =
e^{-\frac{3}{2}H_{0}t} {\cal A}^{\mu}(t,\vec{R})$ equation
(\ref{eq26}) takes the form
\begin{equation}\label{eq28}
\ddot{{\cal A}}^{\mu}-e^{-2H_{0}t}
\nabla _{R}^{2}{\cal A}^{\mu}-\left(\frac{9}{4}H_{0}^{2}
+\alpha\right){\cal A}^{\mu}=0,
\end{equation}
being $\alpha=H^2_0(\kappa^2-2)$ a constant parameter,
$\kappa^2=k^2_{\psi_0}/H^2_0$, and $k_{\psi_0}$ the wavenumber
related to the coordinate $\psi$ on the foliation $\psi=\psi_0$.
Expressing ${\cal A}^{\mu}(t,\vec{R})$ as a Fourier expansion
\begin{equation}\label{eq29}
{\cal A}^{\mu}(t,\vec{R})=\frac{1}{(2\pi)^{3/2}}
\int d^{3}k_{R}\int dk_{\psi} \sum _{\gamma =0}^{3}
\epsilon _{(\gamma)}^{\mu}\left[a_{k_{R}}^{\gamma}e^{i\vec{k}_{R}
\cdot\vec{R}}Q_{k_{R}}(t)+cc\right]\delta(k_{\psi}-\kappa H_0),
\end{equation}
the equation of motion for
the effective 4D electromagnetic modes $Q_{k_{R}}(t)$, becomes
\begin{equation}\label{eq30}
\ddot{Q}_{k_{R}}+\left[k_{R}^{2}e^{-2H_{0}t}
-\left(\frac{9}{4}H_{0}^{2}+\alpha\right)\right]Q_{k_{R}}=0,
\end{equation}
whose general solution is
\begin{equation}\label{eq31}
Q_{k_{R}}(t)=F_{1}{\cal H}_{\nu}^{(1)}[y(t)]+F_{2}{\cal H}_{\nu}^{(2)}[y(t)],
\end{equation}
where $y(t)=(k_{R}/H_{0})e^{-H_{0}t}$, $\nu = (1/ 2 H_0) \sqrt{9
H^2_0 + \alpha}$ and $H_0$ remains constant in a de Sitter
expansion.

The corresponding normalization condition for the modes $Q_{k_{R}}(t)$
becomes
\begin{equation}\label{eq32}
Q_{k_{R}}\dot{Q}_{k_{R}}^{*}-\dot{Q}_{k_{R}}Q_{k_{R}}^{*}=i.
\end{equation}
Therefore, taking into account the Bunch-Davies vacuum condition:
$F_{1}=0$ and $F_2=i \sqrt{(\pi/ 4 H_0)}$, we obtain the solution
of (\ref{eq30})
\begin{equation}\label{eq33}
Q_{k_{R}}(t)=i\sqrt{\frac{\pi}{4H_0}}{\cal H}_{\nu}^{(2)}[y(t)],
\end{equation}
which describes the normalized effective 4D-modes corresponding to
the effective 4D electromagnetic field $A^{\mu}$.
Note that the solution for the 4D-modes of $\chi(t,\vec R) =e^{3 H_0/2}
\varphi(t,\vec R)$ has the same solution as $Q_{k_R}(t)$ in the
equation (\ref{eq33}).

\subsection{Coarse-graining in 4D}

In this section we study the induced effective 4D dynamics of the
fields $\chi (t,\vec{R})$ and
$A^{\beta}(t,\vec{R})$ in a stochastic framework. With this
aim we introduce the corresponding coarse-graining components of 4D redefined scalar and electromagnetic fields
\begin{eqnarray}\label{eq34}
\chi _{L}(t,\vec{R})& = &\frac{1}{(2\pi)^{3/2}}\int d^{3}k_{R}
\int dk_{\psi}\Theta (\vartheta k_{0}-k_{R})\left[a_{k_{R}k_{\psi}}e^{i\vec{k}_{R}
\cdot\vec{R}}\xi _{k_R}(t)+cc\right]\delta(k_{\psi}-\kappa H_0)\\
\label{eq35}{\cal A}_{L}^{\beta}(t,\vec{R})
& = &\frac{1}{(2\pi)^{3/2}}\int d^{3}k_{R}\int dk_{\psi}
\sum_{\lambda = 0}^{3}\epsilon_{(\lambda)}^{\beta}\Theta (\vartheta k_{0}-k_{R})\left[b_{k_{R}k_{\psi}}^{(\lambda)}
e^{i\vec{k}_{R}\cdot\vec{R}}Q_{k_R}(t)+cc\right]\delta(k_{\psi}-\kappa H_0).
\end{eqnarray}
These fields describe the scalar and electromagnetic dynamics in the
IR sector $(k_{R}\ll k_{0})$,
where $k_R/k_0=(k_R/H_0) e^{-H_0 t} < \vartheta \simeq 10^{-3}$. This
implies that in eqs. (\ref{eq34}) and (\ref{eq35})
we are taking into account only modes with wavelenghts larger than $10^3$
times the size of the horizon during inflation.
Analogously, the dynamics in the
short wavelength sector $(k_{R}\gg k_{0})$ is described by the fields
\begin{eqnarray}\label{36}
\chi _{S}(t,\vec{R})&=&\frac{1}{(2\pi)^{3/2}}\int d^{3}k_{R}
\int dk_{\psi}\Theta (k_{R}-\vartheta
k_{0})\left[a_{k_{R}k_{\psi}}e^{i\vec{k}_{R}\cdot\vec{R}}
\xi _{k_{R}}(t)+cc\right]\delta(k_{\psi}-\kappa H_0),\\
\label{eq37}{\cal A}_{S}^{\beta}(t,\vec{R})
&=&\frac{1}{(2\pi)^{3/2}}\int d^{3}k_{R}\int dk_{\psi}
\sum _{\lambda =0}^{3}\epsilon_{(\lambda)}^{\beta}
\Theta (k_{R}-\vartheta k_{0})\left[b_{k_{R}k_{\psi}}
e^{i\vec{k}_{R}\cdot\vec{R}}Q_{k_{R}}(t)+cc\right]
\delta(k_{\psi}-\kappa H_0),
\end{eqnarray}
being $k_{0}^{2}=e^{2H_{0}t}\left(\frac{9}{4}H_{0}^{2}+\alpha\right)$.
As in the previous sections the
relations ${\cal A}^{\beta}={\cal A}^{\beta}_{L}+{\cal A}_{S}^{\beta}$ and
$\chi =\chi _{L} + \chi _{S}$ are satisfied. Thus, its dynamics is governed by
\begin{eqnarray}
\ddot{\chi}_{L}-\omega _{k_R} ^{2}(t)\chi _{L}&=&\vartheta
\left[\ddot{k}_{0}\eta _{3}(t,\vec{R})+\dot{k}_{0}\kappa _{3}(t,\vec{R})
+2 \dot{k}_{0}\gamma _{3}(t,\vec{R})\right], \label{eq38} \\
\ddot{{\cal A}}_{L}^{\beta}-\omega _{k_R} ^{2}(t){\cal A}^{\beta}_{L}
&=&\vartheta\left[\ddot{k}_{0}\eta _{4}^{\beta}(t,\vec{R})+\dot{k}_{0}\kappa _{4}^{\beta}(t,\vec{R})
+2\dot{k}_{0}
\gamma_{4}^{\beta}(t,\vec{R})\right], \label{eq39}
\end{eqnarray}
where $\omega _{k_R}^{2}=e^{-2H_{0}t}(k_{R}^{2}-k_{0}^{2})$ and
\begin{eqnarray}\label{eq40}
\eta_{3}(t,\vec{R})&=&\frac{1}{(2\pi)^{3/2}}\int d^{3}k_{R}\delta(\vartheta 
k_{0}-k_{R})\left[a_{k_{R}}e^{i\vec{k}_{R}\cdot\vec{R}}\xi _{k_{R}}(t)
+cc\right],\\
\label{eq41} \kappa _{3}(t,\vec{R})&=&\frac{1}{(2\pi)^{3/2}}
\int d^{3}k_{R}\dot{\delta}(\vartheta k_{0}-k_{R})\left[a_{k_{R}}e^{i\vec{k}_{R}\cdot\vec{R}}
\xi _{k_{R}}(t)+cc\right],\\
\label{eq42}\gamma _{3}(t,\vec{R})&=&\frac{1}{(2\pi)^{3/2}}
\int d^{3}k_{R}\delta (\vartheta k_{0}-k_{R})\left[a_{k_{R}}e^{i\vec{k}_{R}\cdot\vec{R}}\dot{\xi}_{k_{R}}(t)
+cc\right],\\
\label{eq43} \eta _{4}^{\beta}(t,\vec{R})&=&\frac{1}{(2\pi)^{3/2}}
\int d^{3}k_{R}\sum _{\lambda =0}^{3}
\epsilon_{(\lambda)}^{\beta}\delta (\vartheta k_{0}-k_{R})
\left[b_{k_{R}}^{(\lambda)}e^{i\vec{k}_{R}\cdot\vec{R}}Q_{k_{R}}(t)
+cc\right],\\
\label{eq44} \kappa _{4}^{\beta}(t,\vec{R})
&=&\frac{1}{(2\pi)^{3/2}}\int d^{3}k_{R}\sum _{\lambda =0}^{3}
\epsilon_{(\lambda)}^{\beta}\dot{\delta}(\vartheta k_{0}-k_{R})
\left[b_{k_{R}}^{(\lambda)}e^{i\vec{k}_{R}\cdot\vec{R}}Q_{k_{R}}(t)
+cc\right],\\
\label{eq45} \gamma _{4}^{\beta}(t,\vec{R})
&=&\frac{1}{(2\pi)^{3/2}}\int d^{3}k_{R}\sum _{\lambda =0}^{3}\epsilon _{(\lambda)}^{\beta}\delta (\vartheta k_{0}-k_{R})
\left[b_{k_{R}}^{\lambda}e^{i\vec{k}_{R}\cdot\vec{R}}\dot{Q}_{k_{R}}(t)
+cc\right].
\end{eqnarray}
The second-order stochastic system (\ref{eq38}) and (\ref{eq39}) can
be written as
\begin{eqnarray}\label{eq46}
&&\dot{v}^{\beta}=\omega _{k_R} ^{2}(t){\cal A}_{L}^{\beta}
+\vartheta\dot{k}_{0}\gamma _{4}^{\beta},\qquad \dot{\cal A}_{L}^{\beta}
=v^{\beta}+\vartheta\dot{k}_{0}\eta _{4}^{\beta},\\
\label{eq47} &&\dot{u}=\omega _{k_R}^{2}(t)\chi _{L}
+\vartheta \dot{k}_{0}\gamma _{3},\qquad \dot{\chi}_{L}=
u+\vartheta\dot{k}_{0}\eta _{3},
\end{eqnarray}
with $v^{\beta}=\dot{{\cal A}^{\beta}}-\vartheta\dot{k}_{0}
\eta_{4}^{\beta}$ and $u=\dot{\chi}_{L}-\vartheta\dot{k}_{0}\eta_{3}$.
The conditions to neglect the noise quantities $\gamma _{4}^{\beta}$
and $\gamma _{3}$ compared with $\eta _{4}^{\beta}$
and $\eta _{3}$ respectively, now become
\begin{equation}\label{eq48}
\frac{\dot{Q}_{k_R}\left(\dot{Q}_{k_R}\right)^{*}}{Q_{k_R}Q_{k_R}^{*}} \ll
\left(\frac{\ddot k_0}{\dot k_0}\right)^2,\quad \frac{\dot{\xi}_{k_R}
\left(\dot{\xi}_{k_R}\right)^{*}}{\xi _{k_R}\xi _{k_R}^{*}}
\ll \left(\frac{\ddot k_0}{\dot k_0}\right)^2,
\end{equation}
which are valid on super Hubble scales. The corresponding Fokker-Planck
equations that describe the dynamics of transition probabilities
${\cal P}_{1}\left[\left.({\cal A}_{L}^{\beta})^{(0)},
(v^{\beta})^{(0)}\right|{\cal A}_{L}^{\beta},v^{\beta}\right]$
and ${\cal P}_{2}\left[\left.\chi _{L}^{(0)},u^{(0)}\right|\chi _{L},
u\right]$, are
\begin{eqnarray}\label{eq49}
\frac{\partial {\cal P}_{1}}{\partial t}&=&-v^{\beta}
\frac{\partial {\cal P}_{1}}{\partial {\cal A}_{L}^{\beta}}-
\mu^{2}(t){\cal A}_{L}^{\beta}\frac{\partial {\cal P}_{1}}{\partial
v^{\beta}}+\frac{1}{2}{\cal D}_{11}(t)\frac{\partial ^{2}{
\cal P}_{1}}{\partial (A^{\beta}_L)^{2}},\\
\label{eq50} \frac{\partial {\cal P}_{2}}{\partial t}
&=&-u\frac{\partial {\cal P}_{2}}{\partial \chi _{L}}
-\mu^2(t)\chi _{L}\frac{\partial
{\cal P}_{2}}{\partial u}+\frac{1}{2}
D_{11}(t)\frac{\partial ^{2}{\cal P}_{2}}{\partial \chi _{L}^{2}},
\end{eqnarray}
where $\mu^{2}(t)=e^{-2H_{0}t}k_{0}^{2}(t)$ and the diffusion
coefficients due to stochastic effect of the noises, ${\cal D}_{11}(t)$ and $D_{11}(t)$ related to the variables ${\cal 
A}_{L}^{\beta}$
and $\chi _{L}$, respectively, are
\begin{eqnarray}\label{eq51}
{\cal D}_{11}(t)=\frac{\vartheta ^3}{\pi^2}\dot{k}_{0}k_{0}^{2}
\left|Q_{\vartheta k_{0}}\right|^{2},\\
\label{eq52} D_{11}(t)=\frac{\vartheta ^3}{4\pi ^2}
\dot{k}_{0}k_{0}^{2}\left|\xi _{\vartheta k_{0}}\right|^{2}.
\end{eqnarray}
Hence, the equations of motion for $\left<{\cal A}_{L}^{2}\right> =
\int d{\cal A}_{L}^{\alpha}dv^{\beta}\,({\cal A}_{L})_{\alpha}(
{\cal A}_{L})_{\beta}{\cal P}_{1}\left[{\cal A}_{L}^{\sigma},
v^{\sigma}\right]$ and $<\chi _{L}^{2}>=\int d\chi _{L}du
\chi_{L}^{2}{\cal P}_{2}\left[\chi _{L},u\right]$ are
\begin{eqnarray}\label{eq53}
\frac{d}{d t}\left<{\cal A}_{L}^2\right>&=&\frac{1}{2}{\cal D}_{11}(t)
\simeq
\frac{\vartheta^{3-2\nu}}{8\pi ^3}2^{2\nu}H_{0}^{2\nu}
\Gamma ^{2}(\nu)e^{3H_{0}t}
\left(\frac{9}{4}H_{0}^{2}+\alpha\right)^{\frac{3}{2}-\nu},\\
\label{eq54}\frac{d}{d t}<\chi^2_{L}>
&=&
\frac{1}{2} D_{11}(t) \simeq
\frac{\vartheta^{3-2\nu}}{32\pi ^3}2^{2\nu}H_{0}^{2\nu}
\Gamma ^{2}(\nu)e^{3H_{0}t}
\left(\frac{9}{4}H_{0}^{2}+\alpha\right)^{\frac{3}{2}-\nu}.
\end{eqnarray}
Rewriting expressions (\ref{eq53}) and (\ref{eq54}) in terms of the
original fields $A_{L}^{\beta}(t,\vec{R})=
e^{-\frac{3}{2}H_{0}t}{\cal A}_{L}^{\beta}(t,
\vec{R})$ and $\varphi _{L}=e^{-\frac{3}{2}H_{0}t}\chi _{L}$, we obtain
\begin{eqnarray}\label{eq55}
\frac{d}{d t}<A _{L}^{2}>&=& -3H_{0}<A_{L}^{2}>+
\frac{\vartheta ^{3-2\nu}}{8\pi ^3}2^{2\nu}H_{0}^{2\nu}\Gamma 
^{2}(\nu)\left(\frac{9}{4}H_{0}^{2}+\alpha\right)^{\frac{3}{2}-\nu},\\
\label{eq56}
\frac{d}{d t}<\varphi _{L}^{2}>&=& -3H_{0}<\varphi _{L}^{2}>+
\frac{\vartheta ^{3-2\nu}}{32\pi ^3}2^{2\nu}H_{0}^{2\nu}\Gamma 
^{2}(\nu)\left(\frac{9}{4}H_{0}^{2}+\alpha\right)^{\frac{3}{2}-\nu}.
\end{eqnarray}
Equation (\ref{eq55}) gives information about the dynamics of the
electromagnetic field $A^{\beta}$ on large-scale. However,
although  $<A_{L}^{2}>$ is not an observable, allows to explain
the appearance of  large-scale magnetic fields. In view of this fact
it is convenient to calculate the amplitude of the seed magnetic
field induced from  $<A_{L}^{2}>$. On the other hand, equation (\ref{eq56}) describes the dynamic of  $<\varphi _{L}^{2}>$,
that has been studied in more detail
in \cite{MB1}. The integration of eqs. (\ref{eq55}) and (\ref{eq56})
give us
\begin{equation}\label{nue1}
\left<\varphi^2_L\right> = \frac{1}{4} \left<A^2_L\right> =
\frac{\vartheta ^{3-2\nu} 2^{2\nu} H_{0}^{2\nu-1}\Gamma
^{2}(\nu)\left(\frac{9}{4}H_{0}^{2}+\alpha\right)^{\frac{3}{2}
-\nu}}{96 \pi^2} + C e^{-3 H_0 t},
\end{equation}
where $C$ is an integration constant.
The power spectrum ${\cal P}(k_R)$ for
$\left<\varphi^2_L\right> = \frac{1}{4} \left<A^2_L\right>=
{e^{-3 H_0 t}\over (2\pi)^3} {\large\int}^{\vartheta k_0}_{0} d^3k_R
\xi_{k_R}(t) \xi^*_{k_R}(t) \sim {\large\int}^{\vartheta k_0}_{0}
{dk_R \over k_R} {\cal P}(k_R)$, is
\begin{equation}\label{nue2}
{\cal P}(k_R) \sim k^{3-2\nu}_R,
\end{equation}
where $k_R$ is the wavenumber related to $R$.
This spectrum is nearly scale invariant for $\nu \simeq 3/2$ (i.e., for
$|\alpha|/H^2_0 \ll 1$). Note that ${\cal P}(k_R)$ is the power
spectrum for both, $\left<\varphi^2_L\right>$ and $\left<
A^2_L\right>$. This is an interesting result, because the spectrum
of $\left<\varphi^2_L\right>$ is determinant for the structure formation
on cosmological scales after inflation.

\section{Large-scale seed magnetic fields}

Once we know the effective
3D spatial components of the electromagnetic potential and
their evolution on cosmological scales, we can calculate
the components of the magnetic field. In this section
we develop a stochastic treatment for this field
on cosmological scales.

\subsection{Induced Seed Magnetic Fields.}

By means of the use of the equations (\ref{eq26}) and (\ref{eq28}) we obtain that the dynamics of the electromagnetic potential 
$A^{i}$ satisfies
\begin{equation}\label{smf1}
\ddot{A}^{i}+3H_{0}\dot{A}^{i}
-e^{-2H_{0}t}\nabla _{R}^{2}A^{i}-\alpha A^{i}=0.
\end{equation}
Considering the physical components of $\vec{A}$ and $\vec{B}$
measured in a comoving frame. Using $\vec\nabla _{R}\cdot
\vec{B}_{com}=0$ and $\vec B_{com} = \vec{\nabla}_R \times
\vec{A}_{com}$ in spherical coordinates\cite{grav1}, equation
(\ref{smf1}) becomes
\begin{equation}\label{smf4}
\ddot{B}^{i}_{com}+H_{0}\dot{B}^{i}_{com}-e^{-2H_{0}t}
\nabla _{R}^{2}B^{i}_{com}
-(\alpha +2H_{0}^{2})B^{i}_{com}=0.
\end{equation}
This expression describes the dynamics of the comoving components of the
seed magnetic field.
Transforming $B^{i}$ according to $B^{i}_{com}(t,\vec{R})
=e^{-\frac{1}{2}H_{0}t}{\cal B}^{i}_{com}$, from equation (\ref{smf4}) we have
\begin{equation}\label{mag1}
\ddot{{\cal B}^{i}}_{com}-e^{-\frac{1}{2}H_{0}t}
\nabla _{R}^{2}{\cal B}^{i}_{com}-\left(\alpha +
\frac{9}{4}H_{0}^{2}\right){\cal B}^{i}_{com}=0
\end{equation}
As in the case of $A^{\mu}$, we can express these components
 as a Fourier expansion
\begin{equation}\label{smf5}
{\cal B}^{i}_{com}(t,\vec{R})=\frac{1}{(2\pi)^{3/2}}
\int d^{3}k_{R}\sum _{l=1}^{3}\epsilon _{(l)}^{i}(k_{R})
\left[b_{k_{R}}^{(l)}e^{i\vec{k}_{R}
\cdot\vec{R}}G_{k_{R}}(t)
+b_{k_{R}}^{(l)\,\dagger}e^{-i\vec{k}_{R}\cdot\vec{R
}}G_{k_{R}}^{*}(t)\right],
\end{equation}
where $b_{k_{R}}^{(l)\,\dagger}$ and $b_{k_{R}}^{(l)}$ are the creation and
annihilation operators and $\epsilon _{(l)}^{i}(k_{R})$ are
the 3-polarisation vectors which satisfy $\epsilon _{(i)}\cdot
\epsilon _{(j)}=g_{ij}$.
Therefore, the equation of motion for
$G_{k_{R}}(t)$ obtained from (\ref{smf4}), acquires the form
\begin{equation}\label{smf7}
\ddot{G}_{k_{R}}+\left[k_{R}^{2}e^{-2H_{0}t}
-\left(\frac{9}{4}H_{0}^{2}+\alpha\right)\right]G_{k_{R}}=0.
\end{equation}
The normalized solution of (\ref{smf7}) is
\begin{equation}\label{smf9}
G_{k_{R}}(t)=i\sqrt{\frac{\pi H_0}{4}}{\cal H}_{\lambda}^{(2)}[w(t)],
\end{equation}
where $\lambda = {1\over 2 H_0} \sqrt{9H^2_0 + 4\alpha}$ and
$\omega(t) = {k_R\over H_0} e^{-H_0 t}$.

\subsection{Coarse-graining treatment for a seed magnetic field}

Now we are able to obtain the induced seed magnetic
field on large-scale related with the electromagnetic potential $A_{L}^{\beta}$ in an analogously manner as we proceeded in the
preview sections. Therefore,
we introduce the 3D coarse-graining field associated with the
redefined components of the magnetic field
${\cal B}^{i}_{com}(t,\vec{R})$ as
\begin{equation}\label{eq57}
\left.{\cal B}^{i}_{L}\right|_{com}(t,\vec{R})=\frac{1}{(2\pi)^{3/2}}
\int d^{3}k_{R}\sum _{l=1}^{3}\epsilon _{(l)}^{i}\Theta (\vartheta 
k_{0}-k_{R})\left[b_{k_R}^{(l)}e^{i\vec{k}_{R}\cdot\vec{R}}G_{k_{R}}(t)
+cc\right],
\end{equation}
where the modes with $k_{R}/k_{0}\ll
\vartheta$ are referred as the outside of the horizon. The short
wave length modes are described by the field
\begin{equation}\label{eq58}
\left.{\cal B}_{S}^{i}\right|_{com}(t,\vec{R})
=\frac{1}{(2\pi)^{3/2}}\int d^{3}k_{R}\sum _{l
=1}^{3}\epsilon _{(l)}^{i}\Theta (k_{R}-\vartheta k_{0})\left[b_{k_R}^{(l)}e^{i\vec{k}_{R}\cdot\vec{R}}G_{k_{R}}(t)+cc\right],
\end{equation}
such that the relation ${\cal B}^{i}_{com}
=\left.{\cal B}_{L}^{i}\right|_{com}+\left.{\cal B}_{S}^{i}\right|_{com}$ is
satisfied. The stochastic equation of motion for $\left.{\cal B}_{L}^{i}\right|_{com}$ is given by
\begin{equation}\label{eq59}
\left.\ddot{{\cal B}^{i}_{L}}\right|_{com}
-\omega _{k_R}^{2}(t)\left.{\cal B}_{L}^{i}\right|_{com}
=\vartheta\left[\ddot{k}_{0}\eta _{5}^{i}(t,\vec{R})+\dot{k}_{0}\kappa _{5}^{i}(t,\vec{R})+2\dot{k}_{0}\gamma 
_{5}^{i}(t,\vec{R})\right],
\end{equation}
being the stochastic operators
\begin{eqnarray}\label{eq60}
\eta _{5}^{i}(t,\vec{R})&=&\frac{1}{(2\pi)^{3/2}}\int d^{3}k_{R}
\sum _{l=1}^{3}\epsilon _{(l)}^{i}\delta (\vartheta k_{0}-k_R)\left[b_{k_R}^{(l)}e^{i\vec{k}_{R}\cdot\vec{R}}G_{k_{R}}(t)
+cc\right],\\
\label{eq61} \kappa _{5}^{i}(t,\vec{R})
&=&\frac{1}{(2\pi)^{3/2}}\int d^{3}k_{R}\sum _{l=1}^{3}\epsilon _{(l)}^{i}\dot{\delta}(\vartheta k_{0}-k_{R})\left[b_{k_R}^{(l)}
e^{i\vec{k}_{R}\cdot\vec{R}}G_{k_{R}}(t)+cc\right],\\
\label{eq62} \gamma _{5}^{i}(t,\vec{R})&=&\frac{1}{(2\pi)^{3/2}}
\int d^{3}k_{R}\sum _{l=1}^{3}\epsilon _{(l)}^{i}\delta(\vartheta k_{0}-k_{R})\left[b_{k_R}^{(l)}e^{i\vec{k}_{R}\cdot\vec{R}}
\dot{G}_{k_{R}}(t)+cc\right].
\end{eqnarray}
The equation (\ref{eq59}) can be expressed by the system
\begin{eqnarray}\label{eq63}
\left.\dot{{\cal B}_{L}^{i}}\right|_{com}=W^{i}+\vartheta \dot{k}_{0}
\eta _{5}^{i},\qquad \dot{W^{i}}=\omega _{k_R} ^{2}(t)\left.{\cal B}_{L}^{i}\right|_{com}+\dot{k}_{0}\gamma _{5}^{i},
\end{eqnarray}
where $W^{i}$ is an auxiliary field defined by $W^{i}
=\left.\dot{{\cal B}_{L}^{i}}\right|_{com}
-\vartheta\dot{k}_{0}\eta _{5}^{i}$.
In this system the effect of the noise $\gamma _{5}^{i}$ can
be minimized if $(\dot{k_{0}})^{2}\left<\left(\gamma^i_{5}\right)^{2}\right>
\ll (\ddot{k_{0}})^{2}\left<\left(\eta^i_{5}\right)^{2}\right>$,
which is valid if the condition
\begin{equation}\label{eq64}
\frac{\dot{G}_{k_{R}}\dot{G}_{k_{R}}^{*}}{G_{k_R}G_{k_R}^{*}}\ll
\left(\frac{\ddot k_0}{\dot k_0}\right)^2=H_{0}^{2},
\end{equation}
is satisfied. For a de Sitter expansion this condition means that
the noise $\gamma _{5}^{i}$ can be neglected on scales $k_{R} \ll e^{H_{0}t}\sqrt{(9/4)H_{0}^{2}+\alpha}$, i.e. on super Hubble 
scales. In such a case the system (\ref{eq63}) can be approximated by
\begin{eqnarray}\label{eq65}
\dot{W^{i}}&=&\mu^{2}(t)\left.{\cal B}_{L}^{i}\right|_{com},\\
\label{eq66}
\left.\dot{{\cal B}_{L}^{i}}\right|_{com}
&=&W^{i}+\vartheta \dot{k_0}\eta _{5}^{i}.
\end{eqnarray}
These are two Langevin equations where the noise $\eta^i_{5}$ satisfies
\begin{equation}\label{eq67}
\left<\eta^i _{5}\right>=0,\qquad \left<\left(\eta^i _{5}
\right)^{2}\right>
=\frac{3}{2\pi^2}\frac{\vartheta k_{0}^{2}}{\dot{k}_{0}}
G_{\vartheta k_{0}}G_{\vartheta k_{0}}^{*}\delta(t-t'),
\end{equation}
which means that the noise $\eta^i_{5}$ is Gaussian and white in
nature. As in the case of ${\cal A}_{L}^{\beta}$, the dynamics
of the transition probability ${\cal P}_{3}\left[\left.
(\left.{\cal B}_{L}^{k})\right|_{com}^{(0)},(W^{k})^{(0)}
\right|\left.{\cal B}_{L}^{k}\right|_{com},W^{k}\right]$
is given through the Fokker-Planck equation
\begin{equation}\label{eq68}
\frac{\partial {\cal P}_{3}}{\partial t}=-W^{k}\frac{\partial
{\cal P}_{3}}{\partial \left.{\cal B}_{L}^{k}\right|_{com}}
-\mu^{2}(t)\left.{\cal B}_{L}^{k}\right|_{com}\frac{\partial
{\cal P}_{3}}{\partial W^{k}}
+\frac{1}{2}\bar{D}_{11}(t)\frac{\partial ^{2} {\cal P}_{3}}{\partial
\left.(\left.{\cal B}_{L}^{k}\right|_{com}\right)^{2}},
\end{equation}
with $\bar{D}_{11}(t)=
\frac{3\vartheta ^{3}}{2\pi^{2}}\dot{k}_{0}k_{0}^{2}
\left|G_{\vartheta k_{0}}\right|^{2}$ being the diffusion coefficient
related to $\left.{\cal B}_{L}^{k}\right|_{com}$. Thus the equation
of motion for $\left<\left(\left.{\cal B}_{L}\right|_{com}\right)^{2}\right>
=\int d\left.{\cal B}_{L}\right|_{com}^{i}dW^{j}\left(\left.{\cal B}_{L}
\right|_{com}\right)_{i}\left(\left.{\cal B}_{L}\right|_{com}\right)_{j}
{\cal P}_{3}\left[\left.{\cal B}_{L}^{k}\right|_{com},W^{k}\right]$ is
\begin{equation}\label{eq69}
\frac{d}{dt}\left<\left(\left.{\cal B}_{L}\right|_{com}\right)^{2}\right>
=\frac{1}{2}\bar{D}_{11}(t)\simeq \frac{3\vartheta^{3-2\nu}}{16
\pi^{3}}2^{2\nu}\Gamma^{2}(\nu)H_{0}^{2\nu +2}e^{3H_{0}t}\left(\frac{9}{4}H_{0}^{2}+\alpha\right)^{3/2-\nu}.
\end{equation}
In terms of the original field $\left.B_{L}^{i}\right|_{com}(t,\vec{R})=
e^{-\frac{1}{2}H_{0}t} \left.{\cal B}_{L}^{i}\right|_{com}
(t,\vec{R})$, equation (\ref{eq69})
becomes
\begin{equation}\label{eq70}
\frac{d}{dt}\left<\left(\left.B_{L}\right|_{com}\right)^{2}\right>
=-H_{0}\left<\left(\left.B_{L}\right|_{com}\right)^{2}\right>+
\frac{3\vartheta^{3-2\nu}}{16\pi^{3}}2^{2\nu}\Gamma^{2}(\nu)
H_{0}^{2\nu+2}e^{2H_{0}t}\left(\frac{9}{4}H_{0}^{2}+\alpha\right)^{3/2-\nu}.
\end{equation}
The general solution of this equation is
\begin{equation}\label{eq71}
\left<\left(\left.B_{L}\right|_{com}\right)^{2}\right>=
\frac{\vartheta^{3-2\nu}}{16\pi^{3}}2^{2\nu}\Gamma^{2}(\nu)H_{0}^{2\nu 
+1}\left(\frac{9}{4}H_{0}^{2}+\alpha\right)^{\frac{3}{2}-\nu}
e^{2H_{0}t}+Ce^{-H_{0}t},
\end{equation}
where C is an integration constant. For $\nu =3/2$, this
expression is reduced to
\begin{equation}\label{eq72}
\left<\left(\left.B_{L}\right|_{com}\right)^{2}
\right>=\frac{\Gamma^{2}(\frac{3}{2})}{2
\pi^{3}}H_{0}^{4}e^{2H_{0}t}+Ce^{-H_{0}t}.
\end{equation}

On the other hand the physical magnetic field $B_{phys}$ and
the comoving one are
related by \cite{grav1}
\begin{equation}\label{eq73}
B_{phys}\sim a^{-2}B_{com}.
\end{equation}
After inflation, $B_{phys}$ decreases as $a^{-2}$. Therefore, we could make
an estimation of the actual strength of the cosmological magnetic field $B_{phys}^{(a)}$. Hence, we can use
the expression
\begin{equation}\label{eq74}
\left<\left(\left.B^{(a)}_L\right|_{phys}\right)^{2}\right>^{1/2}_{IR}
\simeq\left(\frac{a(t=t_{0})}{a(t=t_{i})}\right)^{-4}
\left<\left.B^{2}_L\right|_{com}(t=t_{i})\right>^{1/2}_{IR},
\end{equation}
being $t_{i}$ and $t_{0}$ the time at the end of inflation and
 the actual, respectively. We estimate the factor
\begin{equation}\label{eq75}
\left(\frac{a(t=t_{0})}{a(t=t_{i})}\right)^{-4}\simeq 10^{-136},
\end{equation}
where we have used $H_{0}=0.5\times 10^{-9} M_{p}$ that takes
into account $N_{e}=63$ at the end of inflation for $t_i=1.26 \times
10^{11} \  {\rm G}^{1/2}$.
Note that the eq. (\ref{eq75}) accounts for the actual (at $t=t_0$)
size of the observable horizon ($\sim 10^{28}$ cm.) and the size
of the horizon ($\sim 3.6 \times 10^{-6}$ cm.) at the end of inflation
(at $t=t_i$). In order to obtain an estimation of the
actual strength of the magnetic
field, we substitute equation (\ref{eq72}) into (\ref{eq74}) such that
the strength of the magnetic field is
given by
\begin{equation}\label{eq76}
\left.\left<\left(\left.B_{L}^{(a)}\right|_{phys}\right)^2\right>^{1/2}
\right|_{IR}=(4.9448\times 10^{-16})
\left[\frac{\Gamma^{2}(3/2)}{2\pi^{3}}H_{0}^{4}e^{2H_{0}t}
+Ce^{-H_{0}t}\right]^{1/2} \  {\rm Gauss}.
\end{equation}
For an integration constant $C \le 10^{45}$, we obtain
$\left<\left(\left.B_{L}^{(a)}\right|_{phys}\right)^2\right>^{1/2}
\le 10^{-9}$ Gauss, that agrees with some other calculations of cosmological magnetic fields strengths made in \cite{mar}.
Values considered for $\vartheta$ correspond to actual scales from
$3\times 10^{3}$ to $3\times 10^{6}$ {\rm Mpc} and a nearly
scale invariant power spectrum.\\

\section{Final Remmarks}

In this work we have studied the emergence of a classical behavior
from the quantum dynamics of the components of the potential
vector $A_{B} = (A_{\mu},\varphi)$, in gravitoelectromagnetic
inflation, by considering a coarse-grained average. Our approach
leads to a suitable formalism for studying the temporal evolution
of $A_B$ beyond the slow-roll approximation by assuming spatial
homogeneity. Thus, we neglect the gradient term in the equation of
motion when we consider a coarse-grained representation on
cosmological scales. This assumption allows us to develop a rather
simple description of its temporal evolution in terms of two
two-dimensional Fokker-Planck equations (one related to $\chi_L$,
which is the redefined field of $\varphi_L$, and the another
related to ${\cal A}^{\mu}$, which is the redefined field of
$A^{\mu}$), but at the same time, it restricts the cases where the
formalism is applicable. The approach is based on a consistent
semiclassical expansion for $A_B$. In this framework the inflation
in a 4D de Sitter expansion is driven by the vacuum mean value
of the components $A_B$ (whose
only non-null part is $\left<A_5\right>=\varphi_b$), whereas the
long-wavelength modes of the quantum fluctuations reduce to a
quantum system subject to quantum noises originated by the
short-wavelength sector.
In other words, $\left<A^{\mu}\right>$
are considered as null in the universe on an effective de Sitter
expansion, being $A^{\mu}(t,\vec R)$
their space-time fluctuations. On the other hand, $\left<
A_4\right>=\varphi_b$
(which is a constant of $t$ in a 4D de Sitter expansion), is the solution
of $\varphi(t,\vec R)$ on the background (spatially isotropic and homogeneous)
4D metric (\ref{eq25}).
The range of applicability of this
assumption must be carefully considered because the regime of
temporal evolution and the development of spatial inhomogeneities
are related. For an effective 4D de Sitter expansion here studied,
the scales of viability of our approach is $k_R \ll e^{H_0 t}
\sqrt{ (9/4) H^2_0 + \alpha}$, which describes super Hubble
wavelengths during the inflationary expansion. Hence, the system
can be considered as classical on cosmological scales due to the
contribution of the noises $\gamma^i_5$ can be neglected with
respect to $\eta^i_5$.

Finally, we have made an estimation of
$\left.\left<\left(\left.B_{L}^{(a)}\right|_{phys}\right)^2\right>^{1/2}
\right|_{IR}$ for
$\nu\simeq 3/2$ and we obtained values of the order of $ \le
10^{-9}$ Gauss
for scales no much smaller than the actual horizon. Our
results agree with the WMAP CMB data constrains for
cosmological magnetic fields\cite{mar}.
It is remarkable that the results here obtained also agree with other
recently obtained by a different method\cite{grav1}. However,
the advantage of our stochastic method is that the problem
of the divergence for a scale invariant power spectrum (with $\nu=3/2$)
in $\left.\left<\left(\left.B_{L}^{(a)}\right|_{phys}\right)^2\right>^{1/2}
\right|_{IR}$
now is avoided.

\vskip .2cm
\centerline{\bf{Acknowledgements}}
\vskip .2cm
JEMA acknowledges CNPq-CLAF and UFPB
and MB acknowledges CONICET
and UNMdP (Argentina) for financial support.\\

\end{document}